\documentclass[epj]{svjour}
\usepackage{graphicx}

\begin{document}

\title{Zitterbewegung, chirality, and minimal conductivity in graphene}
\titlerunning{Minimal conductivity in graphene}
\author{M. I. Katsnelson}
\institute{Institute for Molecules and Materials, Radboud
University Nijmegen, 6525 ED Nijmegen, The Netherlands}
\date{Received: date / Revised version: date}

\abstract {It has been recently demonstrated experimentally that
graphene, or single-layer carbon, is a gapless semiconductor with
massless Dirac energy spectrum. A finite conductivity per channel
of order of $e^{2}/h$ in the limit of zero temperature and zero
charge carrier density is one of the striking features of this
system. Here we analyze this peculiarity based on the Kubo and
Landauer formulas. The appearance of a finite conductivity without
scattering is shown to be a characteristic property of Dirac
chiral fermions in two dimensions.
}
\maketitle

Graphene, or single layer carbon \cite{kostya1}, demonstrates
unique electronic properties. It has been shown recently
\cite{kostya2,kim} that the charge carriers in graphene are
massless Dirac fermions with effective ``velocity of light'' of
order of 10$^{6}$ ms$^{-1}.$ Graphene provides unexpected
connections between condensed matter physics and quantum field
theory; in particular, a new kind of quantum Hall effect observed
in graphene, that is, half-integer quantum Hall effect
\cite{kostya2,kim,gus,cas} can be considered as a consequence of
the famous Atiyah-Singer index theorem \cite{kostya2}. The latter
guarantees the existence of macroscopically large number of chiral
states with zero energy in external magnetic field.

Another amazing property of graphene is the finite minimal
conductivity which is of the order of the conductance quantum
$e^{2}/h$ per valley per spin; it is important to stress that this
is a ``quantization'' of the conductivity rather than of the
conductance \cite{kostya2}. This is not only very interesting
conceptually but also important in light of potential applications
of graphene for ballistic field-effect transistors \cite{kostya1}.
Therefore the physical origin of the minimal conductivity is worth
the special consideration which is a subject of this Note.

Numerous considerations of the conductivity of a two-dimensional
massless Dirac fermion gas do give this value of the minimal
conductivity with the accuracy of some factor of order of unity
\cite{D1,D2,ludwig,D3,D4,shon,gorbar,D5}. It is really surprising
that in this case there is a final conductivity for an
\textit{ideal} crystal, that is, without any scattering processes
\cite{ludwig}. This fact is important since without complete
understanding of the ideal crystal case one can hardly hope to
have a reliable answer for the realistic case with disorder and
electron-electron interactions. Here we use the Landauer formula
\cite{been} to clarify the physical meaning of this anomaly.

We start with the Hamiltonian of a two-dimensional gapless
semiconductor
\begin{equation}
H=v\sum\limits_{\mathbf{p}}\Psi _{\mathbf{p}}^{\dagger }\mathbf{\sigma p}%
\Psi _{\mathbf{p}}
\end{equation}
and the corresponding expression for the current operator
\cite{abrikos}
\begin{equation}
\mathbf{j}=ev\sum\limits_{\mathbf{p}}\Psi_{\mathbf{p}}^{\dagger
}\mathbf{\sigma }\Psi_{\mathbf{p}} =
\sum\limits_{\mathbf{p}}\mathbf{j}_{\mathbf{p}}
\end{equation}
where $v$ is the electron velocity, $\mathbf{\sigma =}\left(
\sigma _{x},\sigma _{y}\right) $ are Pauli matrices, $\mathbf{p}$
is the momentum, and $\Psi _{\mathbf{p}}^{\dagger }=\left( \psi
_{\mathbf{p}1}^{\dagger },\psi _{\mathbf{p}2}^{\dagger }\right) $
\ are pseudospinor electron operators. Here we omit spin and
valley indices (so, keeping in mind applications to graphene, the
results for the conductivity should be multiplied by 4 due to two
spin projections and two conical points per Brillouine zone).
Straightforward calculations give for the time evolution of the
electron operators
\begin{equation}
\Psi _{\mathbf{p}}\left( t\right) =\frac{1}{2}\left[ e^{-i\epsilon _{\mathbf{%
p}}t}\left( 1+\frac{\mathbf{p\sigma }}{p}\right) +e^{i\epsilon _{\mathbf{p}%
}t}\left( 1-\frac{\mathbf{p\sigma }}{p}\right) \right] \Psi _{\mathbf{p}}
\end{equation}
and for the current operator
\begin{eqnarray}
\mathbf{j}\left( t\right)  &=&\mathbf{j}_{0}\left( t\right) +\mathbf{j}%
_{1}\left( t\right) +\mathbf{j}_{1}^{\dagger }\left( t\right)   \nonumber \\
\mathbf{j}_{0}\left( t\right)  &=&ev\sum\limits_{\mathbf{p}}\Psi _{\mathbf{p}%
}^{\dagger }\frac{\mathbf{p}\left( \mathbf{p\sigma }\right) }{p^{2}}\Psi _{%
\mathbf{p}}  \nonumber \\
\mathbf{j}_{1}\left( t\right)  &=&\frac{ev}{2}\sum\limits_{\mathbf{p}}\Psi _{%
\mathbf{p}}^{\dagger }\left[ \mathbf{\sigma }-\frac{\mathbf{p}\left( \mathbf{%
p\sigma }\right) }{p^{2}}+\frac{i}{p}\mathbf{\sigma \times p}\right] \Psi _{%
\mathbf{p}}e^{2i\epsilon _{\mathbf{p}}t}  \label{current}
\end{eqnarray}
where $\epsilon _{\mathbf{p}}=vp/\hbar$ is the particle frequency.
The last term in Eq.(\ref{current}) corresponds to the
``Zitterbewegung'', a phenomenon connected with the uncertainty of
the position of relativistic quantum particles due to the
inevitable creation of particle-antiparticle pairs at the position
measurement \cite{berest,davydov}. Classical models for this
phenomenon are discussed, e.g., in Ref.\cite{salesi} and
references therein.

In terms of condensed matter physics, the Zitterbewegung is
nothing but a special kind of inter-band transitions with creation
of virtual electron-hole
pairs. The unitary transformation generated by the operator $U_{\mathbf{p}%
}=1/\sqrt{2}(1+i\mathbf{m}_{\mathbf{p}}\mathbf{\sigma )}$, where $\mathbf{m}_{%
\mathbf{p}}=\left( \cos \phi _{\mathbf{p}},-\sin \phi
_{\mathbf{p}}\right) $ and $\phi _{\mathbf{p}}$ is the polar angle
of the vector $\mathbf{p}$, diagonalizes the Hamiltonian
$H_{\mathbf{p}}=diag\left( -vp,vp\right) $ and thus introduces
electron and hole states; after this transformation the
oscillating term in Eq.(\ref{current}) corresponds to the
inter-band transitions, e.g.
\begin{equation}
U_{\mathbf{p}}^{\dagger }j_{\mathbf{p}}^{x}U_{\mathbf{p}}=ev\left(
\begin{array}{cc}
-\cos \phi _{\mathbf{p}} & -i\sin \phi _{\mathbf{p}}e^{-i\phi _{\mathbf{p}%
}+2i\epsilon _{\mathbf{p}}t} \\
i\sin \phi _{\mathbf{p}}e^{i\phi _{\mathbf{p}}-2i\epsilon _{\mathbf{p}}t} &
\cos \phi _{\mathbf{p}}
\end{array}
\right) .  \label{current11}
\end{equation}

To calculate the conductivity $\sigma \left( \omega \right) $ we
will try first to use the Kubo formula \cite{zubarev} which reads
for two-dimensional isotropic case:
\begin{equation}
\sigma \left( \omega \right) =\frac{1}{2A}\int\limits_{0}^{\infty
}dte^{i\omega t}\int\limits_{0}^{\beta }d\lambda \left\langle \mathbf{j}%
\left( t-i\lambda \right) \mathbf{j}\right\rangle  \label{kubo11}
\end{equation}
where $\beta =T^{-1}$ is the inverse temperature, $A$ is the
sample area. In the static limit $\omega =0$ taking into account
Onsager relations and analyticity of the correlators $\left\langle
\mathbf{j}\left( z\right) \mathbf{j}\right\rangle$ for $- \beta <
\mathrm{Im}z \leq 0 $ one has \cite{zubarev}
\begin{equation}
\sigma =\frac{\beta }{4A}\int\limits_{-\infty }^{\infty }dt\left\langle
\mathbf{j}\left( t\right) \mathbf{j}\right\rangle .  \label{kubo}
\end{equation}
Usually, for ideal crystals, the current operator commutes with
the Hamiltonian and thus $\mathbf{j}\left( t\right)$ does not
depend on time. In that case, due to Eq.(\ref{kubo11}) the
frequency-dependent conductivity contains only the Drude peak
\begin{equation}
\sigma _{D}\left( \omega \right) =\frac{\pi }{2A}\lim_{T\rightarrow 0}\frac{%
\left\langle \mathbf{j}^{2}\right\rangle }{T}\delta \left( \omega \right)
\label{Drude}
\end{equation}
Either the spectral weight of the Drude peak is finite and, thus,
the static conductivity is infinite, or it is equal to zero. It is
easy to check that for the system under consideration the spectral
weight of the Drude peak is proportional to the modulus of the
chemical potential $\left| \mu \right| $ (cf. Eq.(44) of
Ref.\cite{D5}) and thus vanishes at zero doping ($\mu =0$). It is
the Zitterbewegung, i.e. the oscillating term
$\mathbf{j}_{1}\left( t\right) $ which is responsible for
nontrivial behavior of the conductivity for zero temperature and
zero chemical potential (that is, the gapless semiconductor case).
A straightforward calculation gives a formal result
\begin{equation}
\sigma =\frac{\pi e^{2}}{2h}\int\limits_{0}^{\infty }d\epsilon \epsilon
\delta ^{2}\left( \epsilon \right)  \label{sigma}
\end{equation}
where one delta-function originates from the integration over $t$
in Eq.(\ref {kubo}) and the second one - from the derivative of
the Fermi distribution function appearing at the calculation of
the average over product of Fermi-operators. Of course, the square
of the delta function is not a well-defined object and thus
Eq.(\ref{sigma}) is meaningless before specification of the way
how one should regularize the delta-functions. After
regularization the integral in Eq.(\ref{sigma}) is finite, but its
value depends on the regularization procedure. It is not
surprising therefore that two different ways of calculations in
Ref.\cite{ludwig} led to two different answers. Our derivation, at
least, clarifies the origin of these difficulties: it is the
Zitterbewegung, or, physically, the impossibility to localize
ultrarelativistic particles and to measure their coordinates.

At finite frequency and finite chemical potential the
Zitterbewegung contribution to the expression (\ref{kubo11})
coincides with the result for inter-band conductivity found in
Ref.\cite{D5}.

Despite this derivation cannot give us a correct numerical factor,
it opens new way to qualitative understanding of more complicated
situations. For example, the minimal conductivity  of order of
$e^{2}/h$ per channel has been observed experimentally also for
the bilayer graphene \cite{natphys} with the energy spectrum
drastically different from that for the single-layer case. The
bilayer graphene is a zero-gap semiconductor with
\textit{parabolic} touching of the electron and hole bands
described by the single-particle Hamiltonian \cite{natphys,falko}
\begin{equation}
H_{\mathbf{p}}=\left(
\begin{array}{cc}
0 & \left( p_{x}-ip_{y}\right) ^{2}/2m \\
\left( p_{x}+ip_{y}\right) ^{2}/2m & 0
\end{array}
\right)   \label{bilayer}
\end{equation}
(here we ignore some complications due to large-scale hopping
processes which are important for a very narrow range of the Fermi
energies \cite{falko}). The Hamiltonian (\ref{bilayer}) can be
diagonalized by the unitary
transformation $U_{\mathbf{p}}$ with the replacement $\phi _{\mathbf{p}%
}\rightarrow 2\phi _{\mathbf{p}}.$ Thus, the current operator after the
transformation takes the form (\ref{current11}) with the replacement $%
v\rightarrow p/m,e^{-i\phi _{\mathbf{p}}}\rightarrow e^{-2i\phi _{\mathbf{p}%
}}$. In contrast with the single-layer case, the density of
electron states for the Hamiltonian (\ref{bilayer}) is finite at
zero energy but the square of the current is, vice versa, linear
in energy. As a result, we have the same estimation (\ref{sigma}),
with the accuracy of additional factor 2.

To circumvent the problem of ambiguity in the expression for
$\sigma$ in Eq.(\ref{sigma}) we now follow the alternative
Landauer approach. Let us assume that our sample is a ring of
length $L_{y}$ in $y$ direction; we will use Landauer formula to
calculate the conductance in $x$ direction (see Fig. 1). There is
still an uncertainty in the sense that the conductivity turns out
to be dependent on the shape of the sample. To have a final
transparency we should keep $L_{x}$ finite. On the other hand,
periodic boundary conditions in $y$ direction are nonphysical and
we have to choose $L_{y}$ as large as possible to weaken their
effects. Thus, for two-dimensional situation one should choose
$L_{x}\ll L_{y}.$

\begin{figure}[tbp]
\includegraphics[width=9cm]{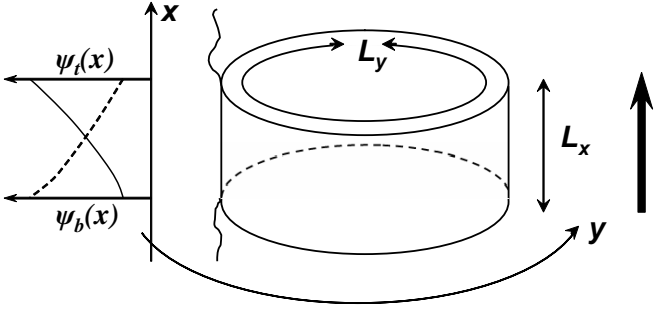}
\caption{Geometry of the sample. Thick arrow shows the direction of current.
$\protect\psi_t$ (solid line) and $\protect\psi_b$ (dashed line) are wave
functions of the edge states localized near the top and the bottom of the
sample, correspondingly. }
\label{fig:1}
\end{figure}

In the coordinate representation the Dirac equation at zero energy takes the
form
\begin{eqnarray}
(K_{x}+iK_{y})\psi _{1} &=&0 \\
(K_{x}-iK_{y})\psi _{2} &=&0  \nonumber
\end{eqnarray}
where $K_{i}=-i\frac{\partial }{\partial x_{i}}.$ General solutions of these
equations are just arbitrary analytical (or complex conjugated analytical)
functions:
\begin{eqnarray}
\psi _{1} &=&\psi _{1}\left( x+iy\right) , \\
\psi _{2} &=&\psi _{2}\left( x-iy\right) .  \nonumber
\end{eqnarray}
Due to periodicity in $y$ direction both wave functions should be
proportional to $\exp \left( ik_{y}y\right) $ where $k_{y}=2\pi
n/L_{y},n=0,\pm 1,\pm 2,....$ This means that the dependence on
$x$ is also fixed: the wave functions are proportional to $\exp
\left( \pm 2\pi nx/L_{y}\right) .$ They correspond to the states
localized near the bottom and top of the sample (see Fig. 1).

To use the Landauer formula, we should introduce boundary
conditions at the sample edges ($x=0$ and $x=L_{x}$). To be
specific, let us assume that the leads are made of doped graphene
with the potential $V_{0}<0$ and the Fermi energy
$E_{F}=vk_{F}=-V_{0}.$ The wave functions in the leads are
supposed to have the same $y$-dependence, that is, $\psi
_{1,2}\left( x,y\right) =\psi _{1,2}\left( x\right) \exp \left(
ik_{y}y\right) .$ Thus, one can try the solution of the Dirac
equation in the following form:
\begin{eqnarray}
\psi _{1}\left( x\right)  &=&\left\{
\begin{array}{cc}
e^{ik_{x}x}+re^{-ik_{x}x}, & x<0 \\
ae^{k_{y}x}, & 0<x<L_{x} \\
te^{ik_{x}x}, & x>L_{x}
\end{array}
\right.   \nonumber \\
\psi _{2}\left( x\right)  &=&\left\{
\begin{array}{cc}
e^{ik_{x}x+i\phi }-re^{-ik_{x}x-i\phi }, & x<0 \\
be^{-k_{y}x}, & 0<x<L_{x} \\
te^{ik_{x}x+i\phi }, & x>L_{x}
\end{array}
\right.   \label{solution}
\end{eqnarray}
where $\sin \phi =k_{y}/k_{F},k_{x}=\sqrt{k_{F}^{2}-k_{y}^{2}}.$
From the conditions of continuity of the wave functions, one can
find the transmission coefficient
\begin{equation}
T_{n}=\left| t\left( k_{y}\right) \right| ^{2}=\frac{\cos ^{2}\phi }{\cosh
^{2}(k_{y}L_{x})-\sin ^{2}\phi }.  \label{T1}
\end{equation}
Further, one should assume that $k_{F}L_{x}\gg 1$ and put $\phi
\simeq 0$ in Eq.(\ref{T1}). Thus, the trace of the transparency
which is just the conductance (in units of $e^{2}/h$) is
\begin{equation}
TrT=\sum\limits_{n=-\infty }^{\infty }\frac{1}{\cosh ^{2}(k_{y}L_{x})}\simeq
\frac{L_{y}}{\pi L_{x}}.  \label{T2}
\end{equation}
Assuming that the conductance is equal to $\sigma
\frac{L_{y}}{L_{x}}$ one finds the contribution to the
conductivity equal to $e^{2}/(\pi h)$. Experimentally
\cite{kostya2}, it is close to $e^{2}/h$, that is, roughly, three
times larger than our estimation. The same result has been found
earlier in Ref.\cite{ludwig} by one of the ways of derivation (the
other one gives, instead, a factor $\pi /8).$ Note also that for
the case of nanotubes ($L_{x}\gg L_{y}$) one has a conductance
$e^{2}/h$ per channel, in accordance with known results
\cite{tian,chico}.

The result $\sigma = e^{2}/(\pi h)$ per valley per spin is found
here for the case of ideal crystal. If one calculates in the
simplest ``bubble'' approximation the conductivity in the presence
of weakly scattering impurities and then put $T=\mu =0$ it leads
to the same value \cite{gus,cas,D2,shon,gorbar}. However, one can
hope that more transparent physical understanding of the origin of
finite conductivity in ideal crystals which is provided by the
Landauer formula will be useful to consider more complicated
situations, such as the case of bilayer \cite{natphys}.

I am thankful to Andreas Ludwig, Andre Geim, and Kostya Novoselov for
valuable discussions stimulating this work.

{\it Note added:} After this work was basically finished
(cond-mat/0512337, revised version) I have become aware of a
relevant work by J. Tworzydlo, B. Trauzettel, M. Titov, A. Rycerz,
and C. W. J. Beenakker (cond-mat/0603315) where a similar result
for the transmission coefficient (\ref{T1}) has been obtained,
with a bit different choice of boundary conditions. They have
found also a sub-Poissonian shot noise in {\it ideal} graphene
similar to that in disordered metals which gives a beautiful
example of the importance of electron Zitterbewegung.

\end{document}